\documentclass[twocolumn,tighten]{aastex631}
\newcommand{\LCO}{\affiliation{Las Cumbres Observatory, 6740 Cortona Drive, Suite 102, Goleta, CA 93117-5575, USA}}
\newcommand{\UCSB}{\affiliation{Department of Physics, University of California, Santa Barbara, CA 93106-9530, USA}}

\newcommand{\UCD}{\affiliation{Department of Physics and Astronomy, University of California, Davis, 1 Shields Avenue, Davis, CA 95616-5270, USA}}

\newcommand{\UT}{\affiliation{Department of Astronomy, University of Texas at Austin, 1 University Station C1400, Austin, TX 78712-0259, USA}}

\newcommand{\CfA}{\affiliation{Center for Astrophysics \textbar{} Harvard \& Smithsonian, 60 Garden Street, Cambridge, MA 02138-1516, USA}}
\newcommand{\UA}{\affiliation{Steward Observatory, University of Arizona, 933 North Cherry Avenue, Tucson, AZ 85721-0065, USA}}

\newcommand{\TAU}{\affiliation{School of Physics and Astronomy, Tel Aviv University, Tel Aviv 69978, Israel}}

\newcommand{\UNC}{\affiliation{Department of Physics and Astronomy, University of North Carolina, 120 East Cameron Avenue, Chapel Hill, NC 27599, USA}}

\newcommand{\JHU}{\affiliation{Department of Physics and Astronomy, The Johns Hopkins University, 3400 North Charles Street, Baltimore, MD 21218, USA}}

\newcommand{\GeminiNorth}{\affiliation{Gemini Observatory, 670 North A`ohoku Place, Hilo, HI 96720-2700, USA}}
\newcommand{\Keck}{\affiliation{W.~M.~Keck Observatory, 65-1120 M\=amalahoa Highway, Kamuela, HI 96743-8431, USA}}

\newcommand{\USask}{\affiliation{Department of Physics \& Engineering Physics, University of Saskatchewan, 116 Science Place, Saskatoon, SK S7N 5E2, Canada}}

\newcommand{\Rutgers}{\affiliation{Department of Physics and Astronomy, Rutgers, the State University of New Jersey,\\136 Frelinghuysen Road, Piscataway, NJ 08854-8019, USA}}

\newcommand{\TAMU}{\affiliation{Department of Physics and Astronomy, Texas A\&M University, 4242 TAMU, College Station, TX 77843, USA}}
\newcommand{\Mitchell}{\affiliation{George P.\ and Cynthia Woods Mitchell Institute for Fundamental Physics \& Astronomy, College Station, TX 77843, USA}}

\newcommand{\IAIFI}{\affiliation{The NSF AI Institute for Artificial Intelligence and Fundamental Interactions, USA}}

\newcommand{\Catalyst}{\altaffiliation{LSSTC Catalyst Fellow}}
\newcommand{\UVa}{\affiliation{Department of Astronomy, 530 McCormick Road, Charlottesville, VA 22904-4325, USA}}
\newcommand{\UTA}{\affiliation{Department of Physics, University of Texas at Arlington, Box 19059, Arlington, TX 76019, USA}}
\newcommand{\Konkoly}{\affiliation{Konkoly Observatory, CSFK, MTA Center of Excellence, Konkoly-Thege M. \'ut 15-17, Budapest, 1121, Hungary}}
\newcommand{\ELTE}{\affiliation{ELTE E\"otv\"os Lor\'and University, Institute of Physics and Astronomy, P\'azm\'any P\'eter s\'et\'any 1/A, Budapest, 1117 Hungary}}
\newcommand{\Szeged}{\affiliation{Department of Experimental Physics, University of Szeged, D\'om t\'er 9, Szeged, 6720, Hungary}}

\usepackage{CJK}

\received{2023 June 9}
\revised{2023 June 27}
\accepted{2023 July 5}
\submitjournal{\textit{The Astrophysical Journal Letters}}

\begin{document}

\title{Shock Cooling and Possible Precursor Emission in the Early Light Curve of the Type~II SN~2023ixf}

\correspondingauthor{Griffin Hosseinzadeh}
\email{griffin0@arizona.edu}

\author[0000-0002-0832-2974]{Griffin Hosseinzadeh}
\UA
\author[0000-0003-4914-5625]{Joseph Farah}
\LCO\UCSB
\author[0000-0002-4022-1874]{Manisha Shrestha}
\UA
\author[0000-0003-4102-380X]{David J.\ Sand}
\UA
\author[0000-0002-7937-6371]{Yize Dong \begin{CJK*}{UTF8}{gbsn}(董一泽)\end{CJK*}\!\!}
\UCD
\author[0000-0001-6272-5507]{Peter J.\ Brown}
\TAMU\Mitchell
\author[0000-0002-4924-444X]{K.\ Azalee Bostroem}
\Catalyst\UA
\author[0000-0001-8818-0795]{Stefano Valenti}
\UCD
\author[0000-0001-8738-6011]{Saurabh W.\ Jha}
\Rutgers
\author[0000-0003-0123-0062]{Jennifer E.\ Andrews}
\GeminiNorth
\author[0000-0001-7090-4898]{Iair Arcavi}
\TAU
\author[0000-0002-6703-805X]{Joshua Haislip}
\UNC
\author[0000-0002-1125-9187]{Daichi Hiramatsu}
\CfA\IAIFI
\author[0000-0003-2744-4755]{Emily Hoang}
\UCD
\author[0000-0003-4253-656X]{D.\ Andrew Howell}
\LCO\UCSB
\author[0000-0003-0549-3281]{Daryl Janzen}
\USask
\author[0000-0001-5754-4007]{Jacob E.\ Jencson}
\JHU
\author[0000-0003-3642-5484]{Vladimir Kouprianov}
\UNC
\author[0000-0001-9589-3793]{Michael Lundquist}
\Keck
\author[0000-0001-5807-7893]{Curtis McCully}
\LCO\UCSB
\author[0000-0002-7015-3446]{Nicolas E.\ Meza Retamal}
\UCD
\author[0000-0001-7132-0333]{Maryam Modjaz}
\UVa
\author[0000-0001-9570-0584]{Megan Newsome}
\LCO\UCSB
\author[0000-0003-0209-9246]{Estefania Padilla Gonzalez}
\LCO\UCSB
\author[0000-0002-0744-0047]{Jeniveve Pearson}
\UA
\author[0000-0002-7472-1279]{Craig Pellegrino}
\LCO\UCSB
\author[0000-0002-7352-7845]{Aravind P. Ravi}
\UTA
\author[0000-0002-5060-3673]{Daniel E.\ Reichart}
\UNC
\author[0000-0001-5510-2424]{Nathan Smith}
\UA
\author[0000-0003-0794-5982]{Giacomo Terreran}
\LCO
\author[0000-0001-8764-7832]{J\'ozsef Vink\'o}
\UT\Konkoly\ELTE\Szeged

\begin{abstract}

We present the densely sampled early light curve of the Type~II supernova (SN) 2023ixf, first observed within hours of explosion in the nearby Pinwheel Galaxy (Messier~101; 6.7~Mpc). Comparing these data to recently updated models of shock-cooling emission, we find that the progenitor likely had a radius of $410 \pm 10\ R_\sun$. Our estimate is model dependent but consistent with a red supergiant. These models provide a good fit to the data starting about 1~day after the explosion, despite the fact that the classification spectrum shows signatures of circumstellar material around SN~2023ixf during that time. Photometry during the first day after the explosion, provided almost entirely by amateur astronomers, does not agree with the shock-cooling models or a simple power-law rise fit to data after 1~day. We consider the possible causes of this discrepancy, including precursor activity from the progenitor star, circumstellar interaction, and emission from the shock before or after it breaks out of the stellar surface. The very low luminosity ($-11\mathrm{\ mag} > M > -14\mathrm{\ mag}$) and short duration of the initial excess lead us to prefer a scenario related to prolonged emission from the SN shock traveling through the progenitor system.

\end{abstract}

\keywords{Circumstellar matter (241), Core-collapse supernovae (304), Red supergiant stars (1375), Stellar mass loss (1613), Supernovae (1668), Type II supernovae (1731)}

\section{Introduction} \label{sec:intro}
Type~II supernovae (SNe~II),\footnote{We use ``SNe~II'' to include both the Type~II-P and II-L photometric subclassifications \citep{barbon_photometric_1979} and exclude the Type~IIb and IIn spectroscopic subclassifications, as well as peculiar events like SN~1987A (see \citealt{arcavi_hydrogen-rich_2016} for a review of hydrogen-rich core-collapse SNe).} the hydrogen-rich explosions of red supergiants (RSGs), offer a unique opportunity to probe the physics of massive-star evolution and the nucleosynthesis of heavy elements. In particular, the last months to years of a massive star's life prior to explosion are poorly understood (see \citealt{smith_mass_2014} for a review). Observational evidence in the form of narrow spectral emission lines from high-ionization states \citep[``flash spectroscopy'';][]{gal-yam_wolf-rayet-like_2014,yaron_confined_2017,bruch_large_2021} and numerical light-curve modeling \citep{morozova_unifying_2017,morozova_measuring_2018} suggest extreme but very brief mass-loss events that produce confined shells of dense circumstellar material (CSM) around some but not all SN~II progenitors (see \citealt{hosseinzadeh_weak_2022}). Several mechanisms for ejecting this mass have been suggested, including nuclear burning instabilities, super-Eddington winds, wave-driven mass loss, and neutrino-enhanced mass loss \citep{yoon_evolution_2010,arnett_turbulent_2011,quataert_wave-driven_2012,shiode_observational_2013,moriya_mass_2014,shiode_setting_2014,smith_preparing_2014,woosley_remarkable_2015,fuller_pre-supernova_2017,wu_diversity_2021}. However, these theories are difficult to test. Because the putative phase of instability is so short lived, we do not observe it in RSGs in the Milky Way, and because of the modest increase in luminosity during this phase relative to the depth of current transient surveys, we do not observe it in other galaxies, with possibly a few exceptions \citep[e.g.,][]{jacobson-galan_final_2022}.

One indirect way to probe the properties of the progenitor is by studying the early SN light curve. In the absence of CSM, the early light curves of core-collapse SNe are powered by shock-cooling emission: energy deposited in the ejecta by the shock breakout that is then radiated away as the ejecta cool (see \citealt{waxman_shock_2017} for a review). This emission has been modeled by \cite{rabinak_early_2011} and \cite{sapir_uv/optical_2017}. Recently, \cite{morag_shock_2023} extended these models to earlier phases and proposed a modified spectral energy distribution (SED) to account for line blanketing in the UV.  These models depend on, among other parameters, the radius of the progenitor star. Therefore, it is possible to estimate the progenitor radius of a core-collapse SN by comparing its early light curve to these models \citep[e.g.,][]{garnavich_shock_2016,rubin_type_2016,arcavi_constraints_2017,piro_numerically_2017,rubin_exploring_2017}. Conversely, in several previous cases, light curves that deviate significantly from the shock-cooling model have been interpreted as indirect evidence of circumstellar interaction (e.g., \citealt{hosseinzadeh_short-lived_2018,andrews_sn_2019,dong_supernova_2021,tartaglia_early_2021,hosseinzadeh_weak_2022,pearson_circumstellar_2023}). For example, an unphysically large progenitor radius derived from shock-cooling modeling might instead be a measurement of the radial extent of the CSM or an extended RSG envelope \citep{hosseinzadeh_weak_2022}.

In this Letter, we present the early light curve of SN~2023ixf, a very nearby SN~II observed within hours of explosion. Its proximity to Earth offers an unprecedented opportunity to test shock-cooling models in detail and probe other unknown physics encoded in its early light curve. \cite{berger_millimeter_2023}, \cite{bostroem_early_2023}, \cite{grefenstette_early_2023}, \cite{jacobson-galan_sn_2023}, \cite{smith_high_2023}, \cite{teja_farultraviolet_2023}, and \cite{yamanaka_bright_2023} analyze the early multiwavelength light curve and spectral series of SN~2023ixf, and a candidate RSG progenitor has been identified by \cite{jencson_luminous_2023}, \cite{kilpatrick_sn_2023}, \cite{pledger_possible_2023}, and \cite{soraisam_sn_2023}. In Section~\ref{sec:obs}, we describe our follow-up campaign, as well as pre-discovery observations posted on the Transient Name Server. In Section~\ref{sec:analysis}, we compare our observed light curve to models of shock-cooling emission and show that the first day of photometry is inconsistent with a smooth rise. In Section~\ref{sec:discuss} we discuss the possible origins of this discrepant emission.  We conclude in Section~\ref{sec:conclude} by outlining the prospects for obtaining such observations in the future.

\section{Observations}\label{sec:obs}
\begin{figure*}
    \centering
    \includegraphics[width=\textwidth]{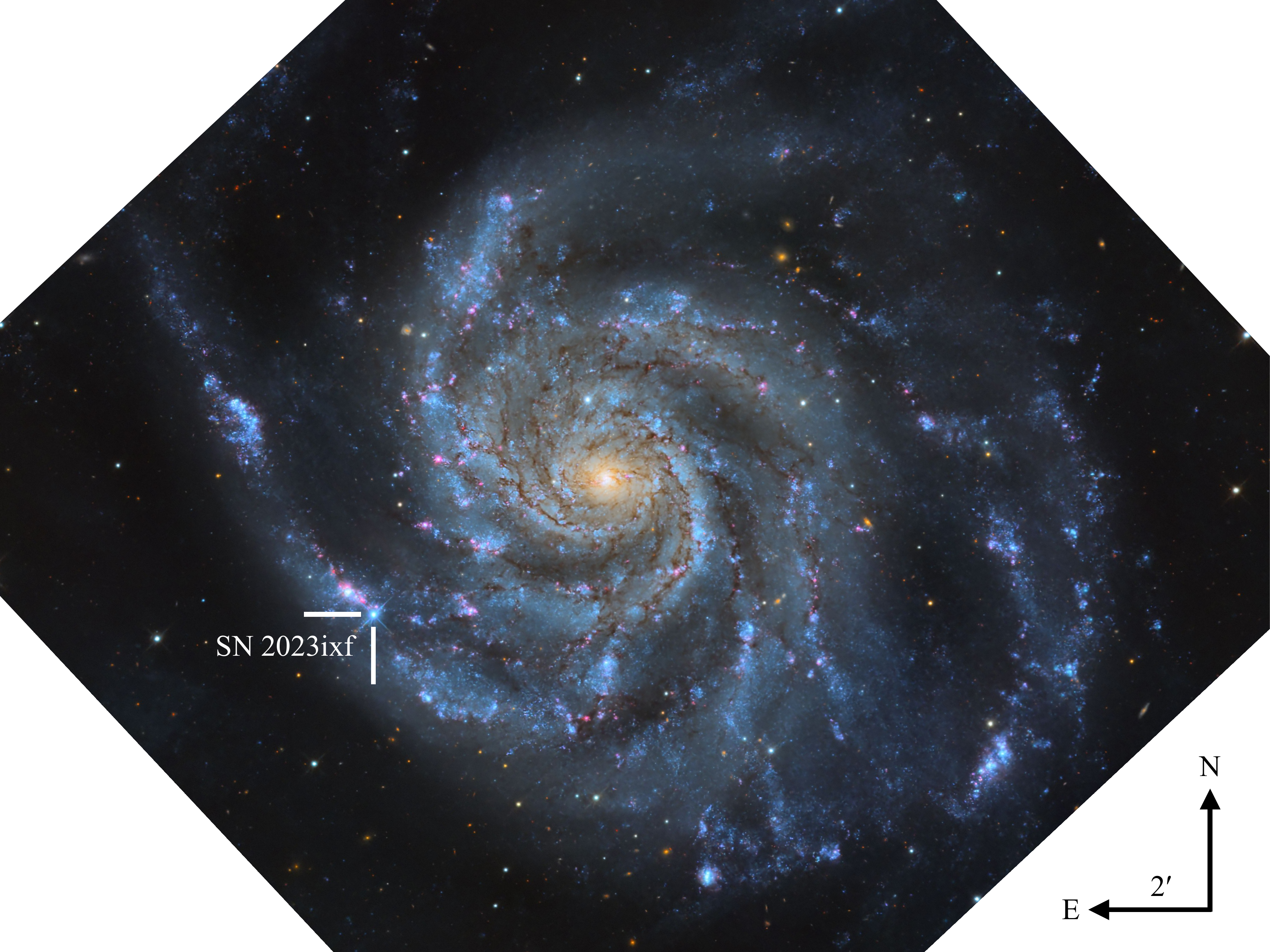}
    \caption{SN~2023ixf occurred in a spiral arm of the Pinwheel Galaxy near several star-forming regions. This image was made using 12~hr of small telescope data on the nights of 2023 May 20, 21, and 22. Image credit: Travis Deyoe, Mount Lemmon SkyCenter, University of Arizona.}
    \label{fig:image}
\end{figure*}

\subsection{Discovery, Classification, and Earlier Detections}
SN~2023ixf was discovered by \cite{itagaki_transient_2023} in an image taken on 2023 May 19 17:27:15 (UT is used throughout; MJD 60083.727) at an unfiltered brightness of 14.9~mag. The SN has J2000 coordinates $\alpha=14\textsuperscript{h}03\textsuperscript{m}38\fs562$, $\delta=+54\degr18'41\farcs94$ \citep{jones_yse_2023}, $264''$ southeast of the center of the Pinwheel Galaxy (Messier~101; NGC~5457; \citealt{evans_chandra_2010}), in a spiral arm near \ion{H}{2} region 39 (Figure~\ref{fig:image}; \citealt{hodge_atlas_1983}). \cite{perley_transient_2023} classified it as Type~II based on a spectrum taken at 22:23:45 on the same day (MJD 60083.933) with the Spectrograph for the Rapid Acquisition of Transients on the Liverpool Telescope (LT) showing strong flash-ionization lines of hydrogen, helium, carbon, and nitrogen (see \citealt{yaron_confined_2017}). \cite{perley_lt_2023} reported additional photometry of $g=13.94$~mag and $r=14.16$~mag from LT images taken immediately before the spectrum (22:09:19; MJD 60083.923).

\begin{figure*}
    \centering
    \includegraphics[width=\textwidth]{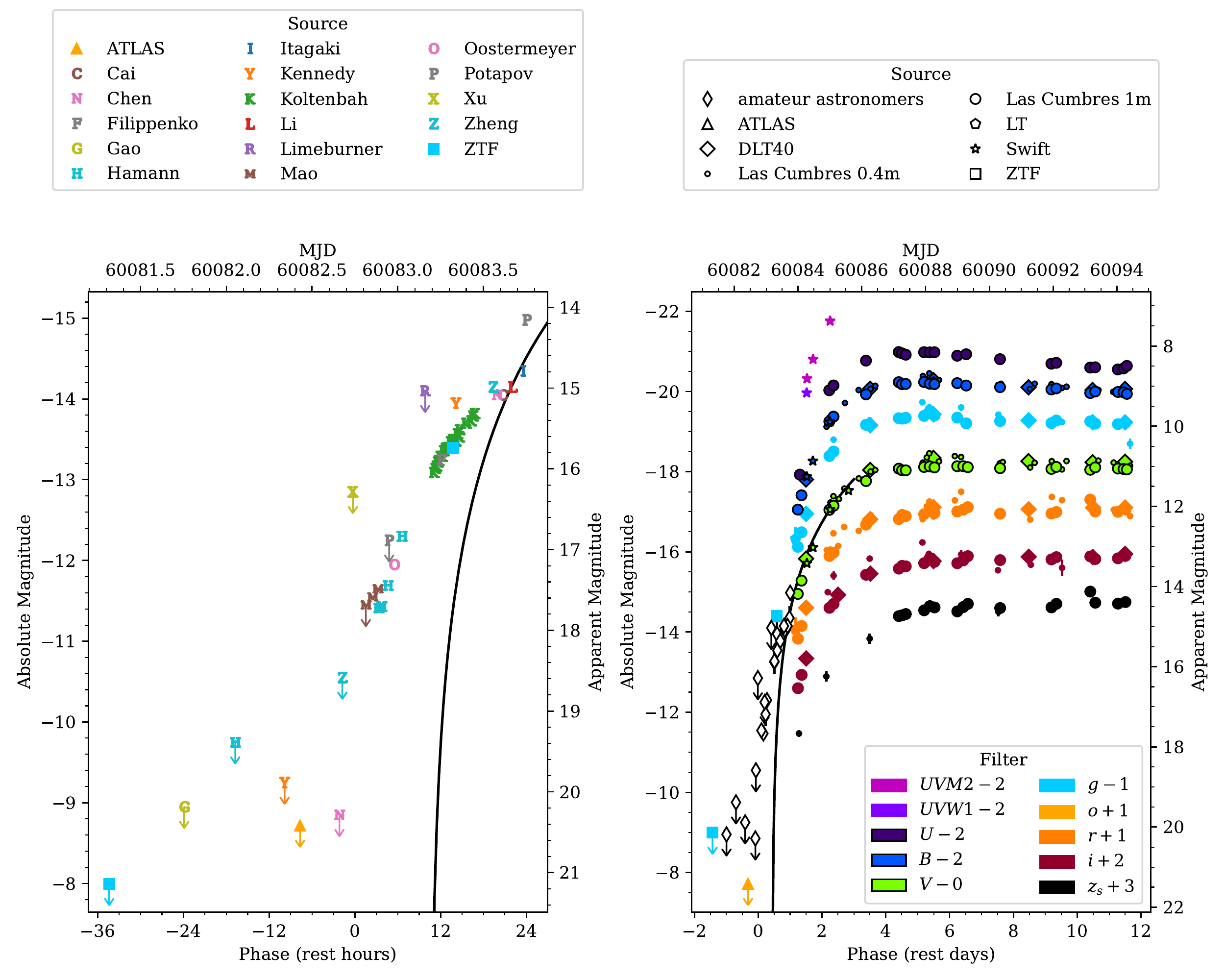}
    \caption{Left: photometry of SN~2023ixf from serendipitous imaging of the Pinwheel Galaxy in the days leading up to discovery by \cite{itagaki_transient_2023}. Apart from the ATLAS ($o$) and ZTF ($g$) points, all the photometry in this panel is derived from unfiltered or color (Bayer RGB) images, and no offset is applied. For clarity, error bars are not plotted, and nondetections are marked with a downward-pointing arrow. The black curve shows an extrapolation of our earliest $V$-band data, assuming $F_\nu \propto t^2$. This extrapolation is in severe disagreement with the data in this panel.
    Right: multiband photometry of SN~2023ixf from Las Cumbres Observatory, the Distance Less Than 40~Mpc (DLT40) Survey, and other sources reported to the Transient Name Server. The white diamonds show the data in the left panel and have no offset applied. The black curve again shows an extrapolation of our earliest $V$-band data, assuming $F_\nu \propto t^2$. This extrapolation is a good description of the early $V$-band data but predicts a much later explosion time than is indicated by the pre-discovery observations. \\
    (The data used to create this figure are available.)}
    \label{fig:phot}
\end{figure*}

After discovery, several amateur and professional astronomers recovered earlier detections and nondetections of SN~2023ixf in their serendipitous observations of the Pinwheel Galaxy.
\cite{perley_ztf_2023} recovered a detection of $g=15.87 \pm 0.01$~mag with the Zwicky Transient Facility (ZTF) 10~hr before discovery and a nondetection of $g > 21.27$ 2 days earlier.
\cite{filippenko_filippenko_2023} reported an unfiltered brightness of $16.0 \pm 0.3$~mag 11~hr before discovery, measured in a JPEG image taken with an 11.4~cm telescope.
\cite{fulton_atlas_2023} reported a nondetection with the Asteroid Terrestrial-impact Last Alert System (ATLAS) of $o > 20.5$~mag 31~hr before discovery.
D.~Kennedy detected the SN using a 15.2~cm telescope at an unfiltered brightness of $\sim$15.3~mag 7--11~hr before discovery and did not detect it 31--35~hr before discovery at ${>}20{-}21$~mag \citep{zhang_prediscovery_2023}.
B.~Oostermeyer detected the SN in a color image taken with a 10~cm telescope at $\sim$17.3~mag 16--20~hr before discovery \citep{zhang_prediscovery_2023}.
\cite{hamann_prediscovery_2023} imaged the SN several times 17--20~hr before discovery with a 30.4~cm telescope and did not detect it the previous night; photometry from his color images was measured by \cite{yaron_amateur_2023}.
\cite{limeburner_limeburner_2023} detected the SN at an unfiltered brightness of $15.5 \pm 0.5$~mag with a 20.3~cm telescope 14~hr discovery.
\cite{mao_onset_2023} report photometry from X.~Gao (60~cm telescope), J.~Chen (20~cm), X.~Zheng (15~cm), S.~Xu (25~cm), Y.~Mao (11~cm), G.~Cai (13~cm), and K.~Li (8~cm) 1--47~hr before discovery.
\cite{koltenbah_prediscovery_2023} detected the SN in a series of 23 images taken 7--12~hr before discovery.
Lastly, N.~Potapov detected the SN with an 8~cm telescope at an unfiltered brightness of $14.27 \pm 0.02$~mag 0.5~hour before discovery and did not detect it to $>$17~mag 20~hr before discovery \citep{chufarin_further_2023}.\footnote{We use the preliminary photometry posted by K.~Sokolovsky on the Transient Name Server: \url{https://www.wis-tns.org/object/2023ixf\#comment-wrapper-38282}.}
Pre-discovery photometry is presented in the left panel of Figure~\ref{fig:phot}.
Where uncertainties are not reported, we conservatively assume 0.3~mag (a ${\sim}3\sigma$ detection) for the purposes of fitting (see Section~\ref{sec:fitting}).
Based on these data, we choose 2023 May 18 18:00 (MJD 60082.75), approximately midway between the first detectionof the SN ($17.7 \pm 0.1$~mag at 20:29 by Mao) and the deep nondetection of Chen ($>$20.4~mag at 15:51), as phase = 0 in our figures.

\subsection{Photometric Follow-up and Data Reduction}
After the discovery and classification reports, we initiated immediate photometric follow-up of SN~2023ixf using the Panchromatic Robotic Optical Monitoring and Polarimetry Telescope \citep{reichart_prompt:_2005} at Sleaford Observatory (Saskatchewan, Canada) as part of the Distance Less Than 40~Mpc (DLT40) Survey \citep{tartaglia_early_2018}; Las Cumbres Observatory's 0.4~m and 1~m telescopes \citep{brown_cumbres_2013} at Teide Observatory (Canary Islands, Spain), McDonald Observatory (Texas, USA), and Haleakal\=a Observatory (Hawai`i, USA) as part of the Global Supernova Project; and the Ultraviolet/Optical Telescope \citep[UVOT;][]{roming_swift_2005} on the Neil Gehrels Swift Observatory \citep{gehrels_swift_2004}. We also include additional observations carried out using Las Cumbres Observatory's 0.4~m telescopes under education and public outreach programs, some of which were first reported by \cite{vannini_amateur_2023,vannini_second_2023,vannini_third_2023}.

\defcitealias{sdsscollaboration_thirteenth_2017}{SDSS Collaboration 2017}

DLT40 data consist of aperture photometry measured using Photutils \citep{bradley_astropy_2022} and calibrated to the AAVSO Photometric All-Sky Survey \citep{henden_aavso_2009}. Las Cumbres data are point-spread function photometry measured using PyRAF-based \texttt{lcogtsnpipe} \citep{valenti_diversity_2016}. \textit{UBV} photometry is calibrated to stars in the SA109 standard field of \cite{landolt_ubvri_1983,landolt_ubvri_1992} observed on the same night with the same telescopes, and $griz_s$ photometry is calibrated to the Sloan Digital Sky Survey \citepalias{sdsscollaboration_thirteenth_2017}.\footnote{We caution that the Las Cumbres $z_s$ filter is significantly different from the Sloan $z$ filter and is more similar to the Pan-STARRS1 $z_s$ filter.} Swift data are reduced following the procedure of the Swift Optical Ultraviolet Supernova Archive (SOUSA; \citealt{brown_sousa:_2014}) and calibrated using the updated zero-points of \cite{breeveld_further_2010} and the time-dependent sensitivity corrections from 2020. Unfortunately, the majority of Swift observations of SN~2023ixf surpassed the maximum count-rate limit for the UVOT detector (see, e.g., \citealt{brown_swift_2012}). Here we use those measurements below that limit using the standard SOUSA analysis. Swift and \textit{UBV} magnitudes are in the Vega system, and $griz_s$ magnitudes are in the AB system. These data are plotted in the right panel of Figure~\ref{fig:phot}.

The Pinwheel Galaxy has a redshift of $z=0.000804 \pm 0.000007$ \citep{devaucouleurs_vizier_1995} and a luminosity distance of $d_L = 6.71 \pm 0.14$~Mpc ($\mu = 29.135 \pm 0.045$~mag; \citealt{riess_determination_2016}), measured via the \cite{leavitt_1777_1908} Law of Cepheid variables. The Milky Way extinction in the direction of SN~2023ixf is $E(B-V)_\mathrm{MW} = 0.0077$~mag \citep{schlafly_measuring_2011}, and the host-galaxy extinction is $E(B-V)_\mathrm{host} = 0.031 \pm 0.006$~mag \citep{lundquist_host_2023,smith_high_2023}. We correct for these using a \cite{fitzpatrick_correcting_1999} extinction law.

\section{Analysis}\label{sec:analysis}
\subsection{The First 24~hr}\label{sec:firstday}
The early light curve of SN~2023ixf consists of a rapid (few-day) rise to a plateau at $M_V \approx -18$~mag. Thanks to the proximity of the host galaxy, we were able to observe this rise starting $\sim$6.5~mag below the plateau. This early phase coverage has almost never been possible for any type of SN \citep[see][]{nugent_supernova_2011,tartaglia_early_2021,jacobson-galan_final_2022}. Notably, Figure~\ref{fig:phot} shows a distinct break in the shape of the rise at a phase of around 24~hr. Before fitting a more physical model, we highlight the break by fitting a simple power law, $F_\nu = A (t - t_0)^2$, with $A$ and $t_0$ as free parameters, to our $V$-band light curve at phases $<3$~days. By adopting $F_\nu(V) \propto t^2$, we assume that the ejecta are in homologous expansion ($R_\mathrm{ej} \propto t$) and that the temperature is constant, the latter of which is certaintly not true at these early times. Nonetheless, this has been a common way of estimating the explosion time of SNe in the literature \citep[e.g.,][]{riess_rise_1999}. The best-fit power law is represented by a black line in both panels of Figure~\ref{fig:phot}. This fit indicates $t_0 \approx 11$~hr, many hours later than the first unfiltered detections around $M \approx -11.5$~mag. The same result holds for other bands.

Our result comes with many caveats. First, the sensitivity functions of the various amateur astronomers' imagers may be significantly different from the transmission function of the $V$ (or any other) filter. However, the pre-discovery detection by ZTF is in excellent agreement with the unfiltered data; it cannot be part of the same $t^2$ rise as the rest of our $g$-band data. Second, the amateur astronomers observed the SN when it was at its faintest and did not perform image subtraction. To measure the background contamination, we downloaded cutouts of the field in $g$, $r$, and $i$ from the Pan-STARRS1 stack images \citep{waters_pan-starrs_2019}, summed the three filters to simulate an unfiltered image, and measured the flux in a $3''$ aperture centered on the SN position using Photutils \citep{bradley_astropy_2022}. We find that the background contributes only 20.87~mag, a $<$6.6\% contaminant in the pre-discovery detections. Lastly, $t^2$ may not be a perfect description of the filtered data, but it is clear that no single power law can describe both the data on days 0--1 and days 1--3. Put simply, the fact that the SN is observed at all $\sim$3~days before plateau cannot be explained by any smooth model. We confirm this result using a physical model in Section~\ref{sec:fitting} and discuss the possible causes in Section~\ref{sec:discuss}.

\subsection{Shock Cooling}\label{sec:fitting}
To further investigate the discrepancy above and to constrain the radius of the progenitor of SN~2023ixf, we compare its early light curve to the shock-cooling model of \cite{morag_shock_2023}. In the shock-cooling paradigm, the star is modeled as a polytrope with a density profile given by $\rho_0 = \frac{3 f_\rho M}{4\pi R^2} \delta^n$, where $f_\rho$ is a numerical factor of order unity, $M$ is the ejecta mass (any remnant is neglected), $R$ is the stellar radius, $\delta \equiv \frac{R-r}{R}$ is the fractional depth from the stellar surface, and $n = \frac{3}{2}$ is the polytropic index for convective envelopes. The product $f_\rho M$ always appears together in the shock-cooling formalism, so we treat it as a single parameter. The shock velocity profile is described by $v_\mathrm{sh} = v_\mathrm{s*} \delta^{-\beta n}$, where $v_\mathrm{s*}$ is a free parameter and $\beta = 0.191$ is a constant. Lastly, $M_\mathrm{env}$ is the mass in the stellar envelope, defined as the region where $\delta \ll 1$. Implicit in their equations is $t_0$, the unknown explosion (core-collapse) time. The approximations made in the model are valid from shock breakout until either recombination power becomes significant or the envelope becomes fully transparent, whichever happens first.

\begin{figure}
    \centering
    \includegraphics[width=\columnwidth]{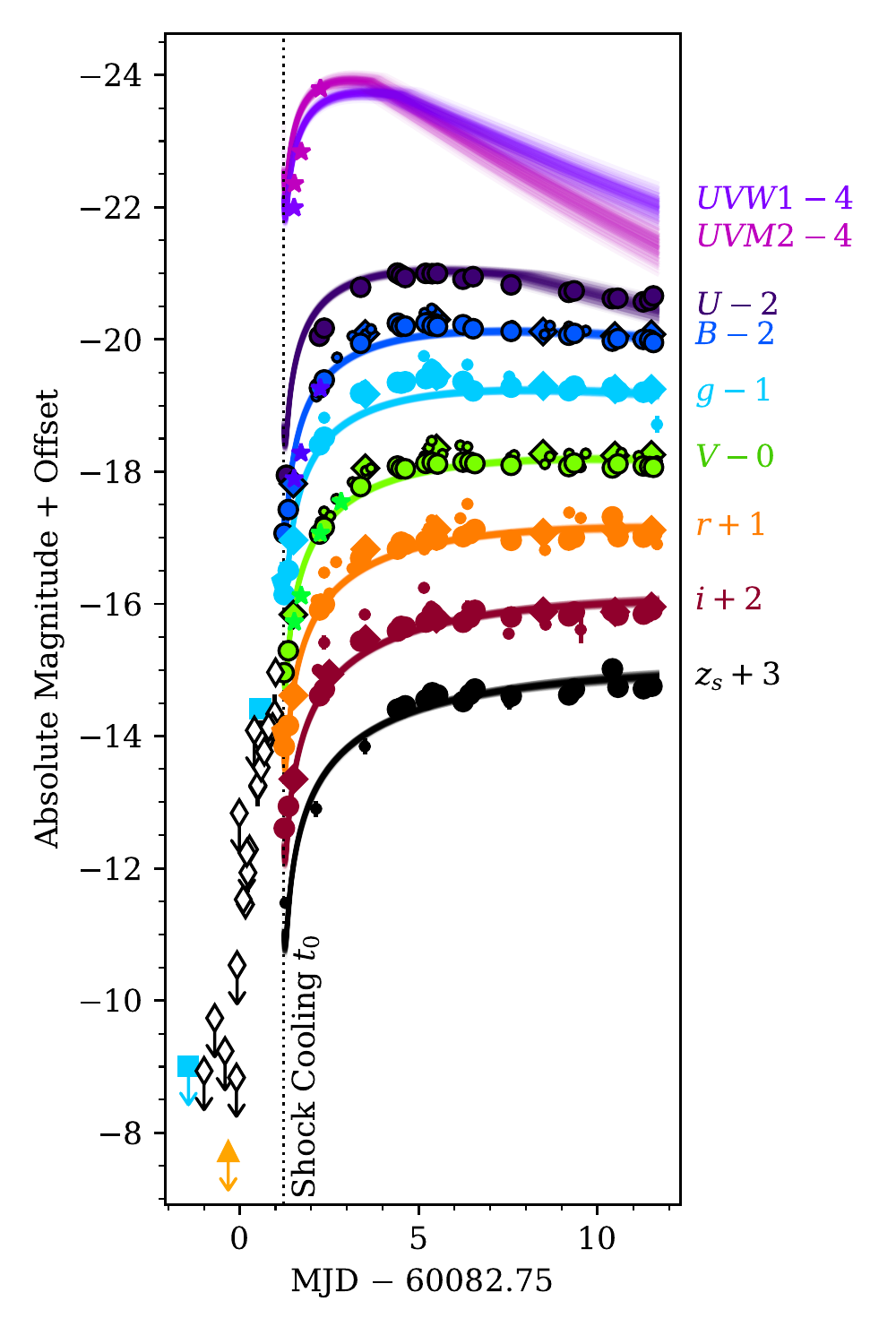}
    \caption{Fitting the shock-cooling model of \cite{morag_shock_2023} to the multiband light curve of SN~2023ixf.
    Markers show the observed photometry, and solid lines are 100 model realizations randomly drawn from the posterior distribution.
    The vertical dotted line shows the best-fit $t_0$ from the shock-cooling model. The SN was detected for nearly a day before this time.}
    \label{fig:fit}
\end{figure}

We implement this model in the Light Curve Fitting package \citep{hosseinzadeh_light_2023a}, and we fit the observed photometry to the model using a Markov Chain Monte Carlo routine. We approximate the sensitivity function for unfiltered data with a top-hat function between 347 and 887~nm (the blue edge of $U$ to the red edge of $I$). For efficiency in fitting, we bin the observed data by 0.1~days. We also include an intrinsic scatter term, $\sigma$, that effectively inflates the observed error bars by a factor $\sqrt{1 + \sigma^2}$. This accounts for scatter around the model as well as potential underestimates of the uncertainties in the photometry. We ran 12 walkers for 3000 steps to reach convergence and an additional 3000 steps to sample the posterior. We confirmed that the best-fit model is valid until $\sim$3 days after the end of the observations presented here, at which point recombination begins to become important. Figure~\ref{fig:fit} shows the results. The fit parameters and their priors and posteriors are described in Table~\ref{tab:params}.

\begin{deluxetable*}{lCcCCCCc}
\tablecaption{Shock-cooling Parameters\label{tab:params}}
\tablehead{&& \multicolumn{3}{c}{Prior} & \multicolumn{2}{c}{Best-fit Values\tablenotemark{a}} & \\[-10pt]
\colhead{Parameter} & \colhead{Variable} & \multicolumn{3}{c}{------------------------------------------} & \multicolumn{2}{c}{------------------------------------} & \colhead{Units} \\[-10pt]
&& \colhead{Shape} & \colhead{Min.} & \colhead{Max.} & \colhead{MSW23} & \colhead{SW17} & }
\startdata
Shock velocity                        & v_\mathrm{s*}  & Uniform & 0      & 10    & 7.2^{+0.5}_{-0.4}        & 3.9^{+0.4}_{-0.2}        & $10^3$ km s$^{-1}$ \\
Envelope mass\tablenotemark{b}        & M_\mathrm{env} & Uniform & 0      & 10    & 5^{+2}_{-1}              & 3.0^{+1.4}_{-0.7}        & $M_\sun$ \\
Ejecta mass $\times$ numerical factor & f_\rho M       & Uniform & 0.3    & 100   & 1.6^{+0.7}_{-0.5}        & 50 \pm 30                & $M_\sun$ \\
Progenitor radius                     & R              & Uniform & 0      & 14374 & 410 \pm 10               & 800^{+200}_{-100}        & $R_\sun$ \\
Explosion time                        & t_0            & Uniform & 82.677 & 84.5  & 83.983^{+0.002}_{-0.003} & 83.868^{+0.006}_{-0.007} & $\mathrm{MJD} - 60000$ \\
Intrinsic scatter                     & \sigma         & Log-uniform & 0  & 10^2  & 12.8 \pm 0.5             & 13.8 \pm 0.6             & \nodata \\
\enddata
\tablenotetext{a}{The ``Best-fit Values'' columns are determined from the 16th, 50th, and 84th percentiles of the posterior distribution, i.e., $\mathrm{median} \pm 1\sigma$. 
MSW23 and SW17 stand for the two models from \cite{morag_shock_2023} and \cite{sapir_uv/optical_2017}, respectively. The former is preferred.}
\tablenotetext{b}{See Section~\ref{sec:fitting} for the definition of ``envelope'' in the shock-cooling paradigm.}
\end{deluxetable*}
\vspace{-24pt}

The model provides a good fit to the multiband post-discovery data. The best-fit progenitor radius is $R = 410 \pm 10\ R_\sun$, consistent with an RSG, though on the small end of the distribution of RSG radii \citep{levesque_astrophysics_2017}. However, as in Section~\ref{sec:firstday}, we find a best-fit explosion time that is inconsistent with the pre-discovery data, in this case at a phase of 1.2~days. We performed another fit that forced the explosion time to be before the first detection by adjusting the upper bound of the prior on $t_0$, but this did not provide a good fit to any part of the light curve. The pre-discovery data were included in both fits, but even so, the model prefers fits with an explosion time after the first detection to accommodate the constraining multiband light curve after $\sim$1~day. We discuss the possible causes of this discrepancy in Section~\ref{sec:discuss}. Because the best-fit model does not pass through any of the pre-discovery points, we conclude that they do not strongly affect any of our parameter estimates.

We also compared our data with an earlier version of the shock-cooling model by \cite{sapir_uv/optical_2017}, which has identical fit parameters but does not account for the very early phase in which the thickness of the emitting shell is smaller than the stellar radius. \cite{morag_shock_2023} built upon this work by interpolating between the treatment of the earlier ``planar'' phase by \cite{sapir_nonrelativistic_2011,sapir_nonrelativistic_2013} and \cite{katz_nonrelativistic_2012} and the treatment of the later ``spherical'' phase by \cite{rabinak_early_2011} and \cite{sapir_uv/optical_2017}. \cite{sapir_uv/optical_2017} also assume a blackbody SED at all times, whereas \cite{morag_shock_2023} account for some line blanketing in the UV.

We repeated our fitting procedure using the model of \cite{sapir_uv/optical_2017}, which is also implemented in the Light Curve Fitting package. The best-fit model looks very similar to the best-fit \cite{morag_shock_2023} model with only a slightly higher intrinsic scatter (13.8 versus 12.8), but the best-fit parameters (Table~\ref{tab:params}) are different. In particular, the parameter $f_\rho M$ has only a weak effect on the light curve in the \cite{sapir_uv/optical_2017} model, and therefore is mostly unconstrained. For this reason, we recommend using the updated model in future studies.

\vspace{1cm}
\section{Discussion}\label{sec:discuss}
In Section~\ref{sec:fitting}, we fit the very early light curve of SN~2023ixf to a recent shock-cooling model to infer parameters about the explosion and progenitor star.  The agreement between our postdiscovery observations and the shock-cooling model is striking, especially given the difficulties in fitting previous versions of this model to other multiband light curves of SNe~II (e.g., \citealt{hosseinzadeh_short-lived_2018,andrews_sn_2019,dong_supernova_2021,tartaglia_early_2021,hosseinzadeh_weak_2022,pearson_circumstellar_2023}). In some of those cases, the model light curves appear to match the data well, but the model parameters are unreasonable for an RSG progenitor. In particular, the best-fit radius is often unphysically large, which might indicate shock breakout in a dense wind or an extended RSG envelope \citep[e.g., ${\sim}2000\ R_\sun$;][]{hosseinzadeh_weak_2022}. However, we have also shown that the estimated progenitor radius is strongly model dependent \citep[see][]{modjaz_shock_2009}, varying by a factor of $\sim$2 between the models of \cite{sapir_uv/optical_2017} and \cite{morag_shock_2023} in the case of SN~2023ixf. In both cases, the best-fit radius is reasonable or even on the low end of the distribution of RSG radii \citep{levesque_astrophysics_2017}, despite the fact that its spectra were dominated by circumstellar interaction during the time period investigated here \citep{bostroem_early_2023,grefenstette_early_2023,jacobson-galan_sn_2023,smith_high_2023,teja_farultraviolet_2023,yamanaka_bright_2023}. It remains to be seen whether future invocations of the model of \cite{morag_shock_2023} can routinely match other early SN light curves, even in the presence of CSM, or whether this result is just a coincidence.

SN~2023ixf is one of the very few SNe where stringent constraints on the progenitor star can be derived both from direct pre-explosion imaging and from the early light curve. The progenitor identified by \cite{kilpatrick_sn_2023} has a luminosity of $\log(L/L_\sun) = 4.74 \pm 0.07$ and a temperature of $T = 3920^{+200}_{-160}$~K. They assume a slightly larger distance of 6.85~Mpc; corrected to our distance, the luminosity is $\log(L/L_\sun) = 4.70 \pm 0.07$. From these values, we derive a progenitor radius of $505^{+67}_{-57}\ R_\sun$, assuming log-normal distributions for $L$ and $T$ and neglecting any covariance. This is consistent at ${\sim}1.5\sigma$ with the result from our shock cooling fitting.

We also found in Section~\ref{sec:analysis} that the pre-discovery photometry is in severe tension with both the shock-cooling model and a simple $t^2$ rise (see Figure~\ref{fig:phot}). We are not aware of any examples of observed excesses in the range $-11\ \mathrm{mag} > M > -14\ \mathrm{mag}$ in the literature. Such data are rarely possible to obtain, especially at the $\sim$hour timescales achieved for SN~2023ixf. Nonetheless, one possibility is that the first day of observations is in fact before core collapse. In this case, the rising light curve we see is due to an eruption or instability in the progenitor itself. For example, RSGs are thought to undergo extreme mass-loss events during the last stages of nuclear burning before core collapse \citep[e.g.,][]{yoon_evolution_2010,arnett_turbulent_2011,quataert_wave-driven_2012,shiode_observational_2013,moriya_mass_2014,shiode_setting_2014,smith_preparing_2014,fuller_pre-supernova_2017,wu_diversity_2021}. These types of mass loss events could explain the precursor activity observed in the RSG progenitor of SN~2020tlf \citep{jacobson-galan_final_2022}, which was around the same magnitude range as in SN~2023ixf but for a much longer period of time ($\sim$130~days leading up to explosion). However, even among these models, most predict a longer phase of excitation than we see in SN~2023ixf.

Another possibility is that core collapse occurred shortly before the first reported detection but that the rising light curve of the SN was strongly affected by the presence of dense CSM, either by increasing the photon diffusion time or by physically slowing the ejecta. Early excesses in some Type~I superluminous SNe have been attributed to CSM interaction \citep[e.g.,][]{leloudas_sn_2012,nicholl_diversity_2015,nicholl_seeing_2016,smith_des14x3taz:_2016,vreeswijk_early-time_2017}. Although these excesses are much more luminous ($M \lesssim -20$~mag) than what we observe in SN~2023ixf, and the progenitors of superluminous SNe and SNe~II are distinct, the mechanism could be related.

In fact, these two possibilities are not independent of each other: the progenitor of SN~2023ixf could have experienced a massive eruption 1 day before explosion, creating a dense shell of material that the ejecta collided with the following day. This scenario is reminiscent of luminous blue variable outbursts and explosions, such as SN~2009ip \citep{smith_discovery_2010,foley_diversity_2011,mauerhan_unprecedented_2013,graham_clues_2014,graham_clues_2017} although again, these outbursts were more luminous ($-14\ \mathrm{mag} > M > -15\ \mathrm{mag}$) and longer lasting ($\sim$30~days) than what we see here, and the progenitors are physically distinct.

Lastly, we may be seeing the shock breakout itself at these very early times \citep[e.g.,][]{schawinski_supernova_2008,garnavich_shock_2016,bersten_surge_2018,forster_delay_2018}, not just the shock-cooling emission. The shock breakout is expected to last only minutes to hours \citep{nakar_early_2010,shussman_type_2016,kozyreva_shock_2020} but can be extended in the presence of CSM by increasing the radius of the photosphere \citep{ofek_supernova_2010,forster_delay_2018}. Recent 3D simulations of SNe~II have also shown that substantial density fluctuations in RSG atmospheres and winds can cause shock breakout at different times across the stellar surface, smearing out the observed signal \citep{goldberg_shock_2022}. Such density fluctuations might be exaggerated in an unstable pre-SN star experiencing strong mass loss, as compared to normal quiescent RSGs. Furthermore, \cite{nakar_early_2010} suggest, and \cite{schawinski_supernova_2008} claim to see, that photons produced in the shock can leak out of the ejecta $\sim$5~hr prior to shock breakout. \cite{garnavich_shock_2016} also see hints of pre-breakout emission in SN light curves from the Kepler Space Telescope but at ${<}3\sigma$ significance. In all these studies, the imaging is not as deep nor the time coverage as dense as for SN~2023ixf, so it is difficult to compare the behavior in detail. However, the very short timescale and low luminosity expected for pre-breakout shock emission make this the most promising explanation for the pre-discovery light curve of SN~2023ixf.

\section{Summary and Conclusions}\label{sec:conclude}
We have presented the early light curve of SN~2023ixf, a very nearby SN~II observed within hours of explosion at a persistent, high cadence over the course of the first two weeks. We find that the first day of photometry is inconsistent with models of shock-cooling emission or a simple $F_\nu \propto t^2$ rise. Various physical mechanisms may be able to explain this excess early emission, in full or in combination, including precursor emission from the RSG progenitor, interaction with CSM, and emission from the core-collapse shock leading up to and including its breakout from the stellar surface. Despite this, the shock-cooling model of \cite{morag_shock_2023} provides a good fit to the data $\gtrsim$1~day after explosion, allowing us to derive a progenitor radius of $R = 410 \pm 10\ R_\sun$, which is consistent with an RSG progenitor although on the small end of the distribution. We show that this estimate of the radius is strongly model dependent, even between different shock-cooling models.

Early SN observations starting at $M > -12$~mag are exceedingly rare due to the limiting magnitudes of most ongoing surveys. The Legacy Survey of Space and Time and Vera C.\ Rubin Observatory will reach these depths on a regular basis for galaxies out to $\sim$100~Mpc \citep{ivezic_lsst:_2019}, potentially revealing previously unseen activity in SN progenitors \citep{jacobson-galan_final_2022}, but its nominal few-day cadence will not be high enough to routinely catch SNe within hours of explosion. Therefore, targeted surveys like DLT40 (but deeper) will still play a crucial role in transient astronomy for the foreseeable future. Likewise, rapid and frequent follow-up from robotic, longitudinally distributed facilities like Las Cumbres Observatory are critical to collecting the dense datasets with which to test and extend theoretical models. Lastly, the incredible success story of SN~2023ixf demonstrates that amateur astronomers can and do contribute invaluable observations to SN science, even in the ``big data'' era of transient astronomy.

SN~2023ixf will be studied for years to come, offering an unprecedented opportunity to understand the physics of the explosion along with the progenitor system and its environment. This range of observations and future theoretical insights may shed light on the early light-curve observations here.

\section*{Acknowledgements}
We are grateful to Travis Deyoe, Alan Strauss, and Mount Lemmon SkyCenter at the University of Arizona for providing the image in Figure~\ref{fig:image}. We also thank Mateo Ballesteros, Fernando C\'ordova, Andr\'es Col\'an Sifuentes, Bruno D\'iaz Yaip\'en, Andr\'e Deback\`ere, Maria Eleftheriou, Dirk Froebrich, Irma Fuentes, Edward Gomez, Jose Guti\'errez, Roger Jim\'enez, Adam Lloyd, Fernando Mujica, Chandru Narayan, Michalis Orfanakis, Karime Osses, Myriam Pajuelo, Aldo Panfichi, Johan Phan, Gabriela Roch, Nayra Rodr\'iguez Eugenio, Valentina Soto, Jasper Swain, Shay Tindol, Julio Vannini, and Romy Weitzer for obtaining some of the Las Cumbres Observatory images analyzed here (programs FTPEPO2014A-004, LCOEPO2014B-010, LCOEPO2018A-001, LCOEPO2018A-005, LCOEPO2018A-009, LCOEPO2019A-005, LCOEPO2020A-001, LCOEPO2021B-007, LCOEPO2022B-003, LCOEPO2022B-008, and LCOEPO2022B-009). Finally, we thank Ido Irani, Jonathan Morag, and Avishay Gal-Yam for calling our attention to a modeling error in an earlier draft of this Letter.

Time-domain research by the University of Arizona team and D.J.S.\ is supported by NSF grants AST-1821987, 1813466, 1908972, \& 2108032, and by the Heising-Simons Foundation under grant \#20201864. 
S.V.\ and the UC Davis time-domain research team acknowledge support by NSF grants AST-2008108.
J.E.A.\ is supported by the international Gemini Observatory, a program of NSF's NOIRLab, which is managed by the Association of Universities for Research in Astronomy (AURA) under a cooperative agreement with the National Science Foundation, on behalf of the Gemini partnership of Argentina, Brazil, Canada, Chile, the Republic of Korea, and the United States of America.
I.A.\ acknowledges support from the European Research Council under the European Union's Horizon 2020 research and innovation program (grant agreement number 852097), from the Israel Science Foundation (grant No.~2752/19), and from the United States--Israel Binational Science Foundation.
M.M.\ acknowledges support in part from ADAP program grant No.~80NSSC22K0486, from the NSF AST-2206657 and from the HST GO program HST-GO-16656.
J.V.\ is supported by the Hungarian National Research, Development and Innovation Office grant K-142534. 

\facilities{ADS, LCOGT (SBIG, Sinistro), Sleaford:PROMPT, NED, Swift (UVOT)}

\defcitealias{astropycollaboration_astropy_2022}{Astropy Collaboration 2022}
\software{
Astrometry.net \citep{lang_astrometry_2010},
Astropy \citepalias{astropycollaboration_astropy_2022},
BANZAI \citep{mccully_lcogt_2018},
\texttt{corner} \citep{foreman-mackey_corner.py:_2016},
\texttt{emcee} \citep{foreman-mackey_emcee_2013},
extinction \citep{barbary_extinction_2016},
HEAsoft \citep{nasaheasarc_heasoft:_2014},
IRAF \citep{nationalopticalastronomyobservatories_iraf:_1999},
\texttt{lcogtsnpipe} \citep{valenti_diversity_2016},
Light Curve Fitting \citep{hosseinzadeh_light_2023a},
Matplotlib \citep{hunter_matplotlib:_2007},
NumPy \citep{oliphant_guide_2006},
Photutils \citep{bradley_astropy_2022},
PyRAF \citep{sciencesoftwarebranchatstsci_pyraf:_2012}
}

\bibliography{zotero_abbrev}
\end{document}